\documentclass[%
 reprint,
 amsmath,amssymb,
 aps,
]{revtex4-1}
\usepackage{eqnarray,amsmath}
\usepackage{graphicx}
\usepackage{dcolumn}
\usepackage{bm}
\usepackage{wasysym}

\newcommand{\rubidium}[0]{$^{85}$Rb }
\begin{document}

\preprint{APS/123-QED}

\title{Eddy current imaging with an atomic radio-frequency magnetometer}

\author{Arne Wickenbrock}
\affiliation{Johannes Gutenberg-Universit{\"a}t Mainz, 55128 Mainz, Germany}
\email{wickenbr@uni-mainz.de}
\author{Nathan Leefer}%
\author{John W. Blanchard}
\affiliation{Helmholtz Institut Mainz, 55099 Mainz, Germany}
\author{Dmitry Budker}
\affiliation{Helmholtz Institut Mainz, 55099 Mainz, Germany,\\ Johannes Gutenberg-Universit{\"a}t  Mainz, 55128 Mainz, Germany,\\ Department of Physics, University of California, Berkeley, CA 94720-7300 and \\ Nuclear Science Division, Lawrence Berkeley National Laboratory, Berkeley, CA 94720}



\date{\today}

\begin{abstract}
We use a radio-frequency \rubidium alkali-vapor cell magnetometer based on a paraffin-coated cell with long spin-coherence time and a small, low-inductance driving coil to create highly resolved conductivity maps of different objects. We resolve sub-mm features in conductive objects, we characterize the frequency response of our technique, and by operating at frequencies up to 250\,kHz we are able to discriminate between differently conductive materials based on the induced response. The method is suited to cover a wide range of driving frequencies and can potentially be used for detecting non-metallic objects with low DC conductivity.

\end{abstract}

\pacs{Valid PACS appear here}
\maketitle


\section{\label{sec1}Introduction}
Magnetic induction measurements 
have many applications in defense, environmental surveying, and in the process and quality control industry~\cite{MIT3,MIT0,Crack1}. The working principle is simple: an oscillating or pulsed magnetic field induces eddy currents in conductive objects and these currents produce a magnetic response that is measured in turn. The specifics of this response depend on intrinsic material properties.
The penetration, or skin depth, of the modulated magnetic field is a function of the conductivity and permeability of the object and the applied frequency.
It has recently been shown that eddy current detection is capable of imaging through metallic enclosures by varying the applied drive frequency and therefore the skin depth~\cite{Ferrucio1}.
The relevance of this in a security context has been pointed out in~\cite{Ferrucio2}. 
The cited works and the majority of commercially available detectors for industry use make use of coils as sensors. Eddy current sensors based on giant magnetoresistance (GMR) magnetometer~\cite{GMR1,GMR2} and super-conducting quantum interference devices (SQUIDs)~\cite{SQUIDS1,SQUID2,SQUID3} have also been implemented.

Atomic magnetometers~\cite{Budker} can reach detection sensitivities approaching SQUID based sensors without the need for a cryogenic environment and they can be miniaturized  for applications requiring compact sensors~\cite{Minis}. The use of atomic magnetometers is beneficial for low and high driving frequencies. At low frequencies, and therefore large skin depth, atomic magnetometers are orders of magnitude more sensitive to oscillating magnetic fields than coil or GMR based sensors. The same image quality can therefore be produced with lower applied radio-frequency (rf) power or images can be acquired through thicker-walled conductive enclosures.
The sensitivity of atomic magnetometers at high frequencies changes little. This allows for biological and bio-medical applications, e.g. creating maps of the saline concentration in biological tissue~\cite{MIT01} or non-contact measurements of the human heart rate~\cite{MedicMIT}.
Furthermore, the ability to create images with the same sensor over a wide range of detection frequencies is beneficial~\cite{Multifrequency1} for the sensitive discrimination of different materials, as demonstrated in this letter.

In \cite{MIT1} an all-optical atomic magnetometer in self-oscillating mode \cite{SelfOsciMag} was used to demonstrate the application of atomic magnetometers for magnetic induction measurements. This technique was then used to create magnetic induction tomographic (MIT) maps of different metallic objects \cite{MIT2}. However, this self-oscillating magnetometer had inherent bandwidth limitations and was less practical because the signal had to be demodulated at the Larmor frequency of the atoms to gain access to the eddy current response. Recently, an atomic magnetometer was used to partially overcome these issues and demonstrate operation at frequencies up to 10\,kHz \cite{MIT4}. 

In this work we use an all-optical atomic magnetometer based on \rubidium that is directly sensitive to the phase and amplitude of a radio-frequency magnetic field that is resonant with the Larmor precession of \rubidium atoms in a static magnetic field. The Larmor frequency can be selected by tuning the static magnetic field at the position of the vapor cell. We are able to create images using effectively arbitrary radio frequencies with this method and present images acquired with up to a 250\,kHz drive-frequency. This is a factor 25 higher than in \cite{MIT2,MIT4} and limited only by the frequency range of the read-out electronics. 

We first present the experimental setup and its characterization; then describe the experimental procedure to perform spatially resolved eddy-current measurements and present the results for a selection of different sample geometries. These include the image of a conductive ring with a 1\,mm slit as an example relevant to crack detection. Finally, we present data of the eddy-current response as a function of the driving frequency for different metals. The latter enables clear material discrimination and demonstrates a promising application of radio-frequency alkali-vapor cell magnetometers.

\section{Experimental setup}\label{sec3}

\begin{figure}[!]
  \centering
  \includegraphics[width=\columnwidth]{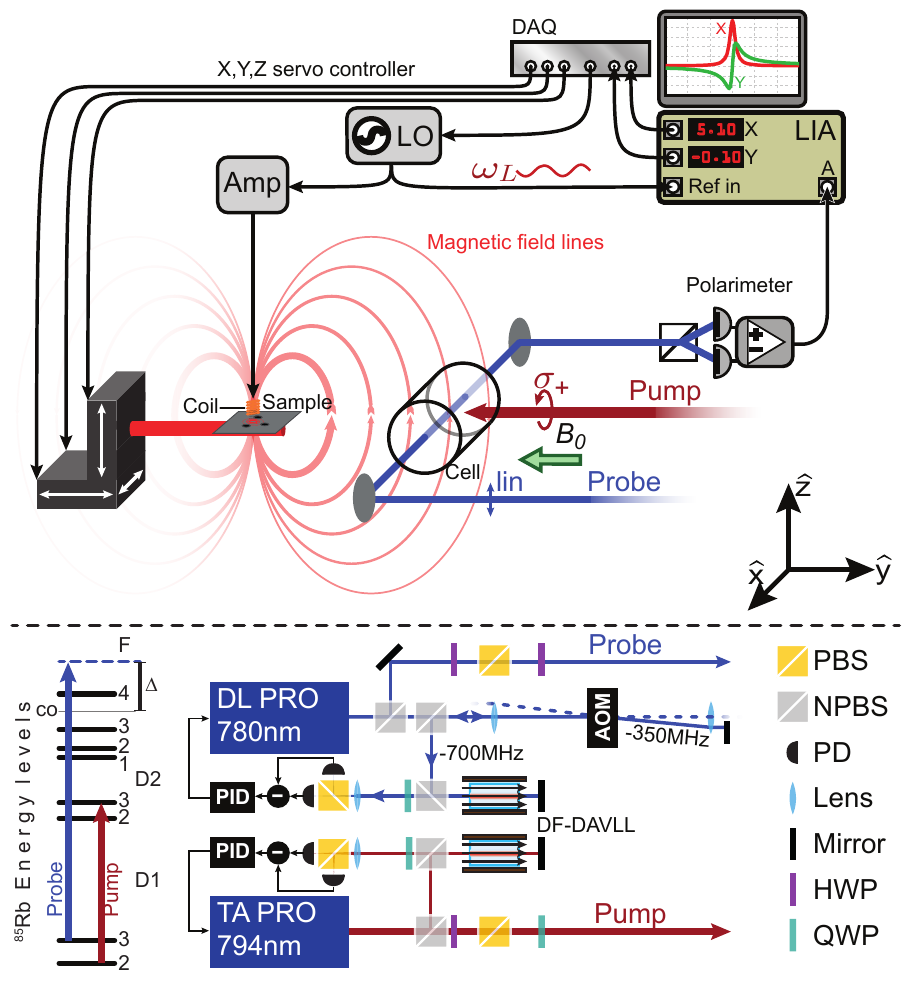}
\caption{Schematic of the experimental setup. The magnetometer consists of a cylindrical paraffin-coated, evacuated vapor cell housed in a cylindrical 4-layer magnetic shield open at one end. A variable magnetic field is applied in y-direction. The circularly polarized pump beam is locked via a DF-DAVLL to the D1 $F=2 \rightarrow F'=3$ transition of \rubidium and propagating parallel to the leading magnetic field. It prepares the atoms in the cell in the $F=3,m_F=3$ eigenstate. Spin precession is driven by applying an oscillating field at the Larmor frequency $\omega_L$ with a small coil oriented in the z-direction and mounted in the plane spanned by the pump and the probe beam. It is shifted 5\,cm in the y-direction from the cell center so that the return flux of the driving coil points predominantly along the z-direction in the cell. The linearly polarized probe beam propagates along the x-direction through the cell center. A small fraction of the probe light is passed twice through a 350\,MHz AOM and then used in a DF-DAVLL to lock the laser to the D2 $F=3 \rightarrow F'=3,4$ crossover feature. The probe beam polarization oscillation is detected with a balanced polarimeter which is read out by a computer controlled dual-phase lock-in amplifier (LIA).
The sample is mounted on a 15\,cm long plastic extension which is attached to a 3D translation stage and moved with respect to the driving coil.}\label{Figure1}
\end{figure}

A schematic of the experimental setup is shown in Fig.~\ref{Figure1}.
At its heart is an evacuated cylindrical paraffin-coated cell with a lockable sidearm (stem) loaded with \rubidium. The cell's outer diameter and length are 25\,mm and it is centered in a four-layer magnetic shield with three outer layers constructed of mu-metal and the innermost layer of ferrite (Twinleaf MS-1F). 
The shield is left open to one side to enable access to the sample.

A circularly polarized laser beam resonant with the \rubidium D1 line, (Toptica TA Pro, 794\,nm, 600\,mW, $\diameter\approx5$\,mm) produces atomic spin polarization along the pump beam propagation axis. The unusually large amount of laser power is used to broaden the magnetic resonance, and therefore reduce the sensitivity to transient magnetic field changes caused by activity in the lab. The laser is frequency stabilized to the D1 \rubidium $F=2 \rightarrow F'=3$ transition with a Doppler-free dichroic atomic vapor laser lock setup (DF-DAVLL) \cite{DAVLL1,DAVLL2,DAVLL3}. 

A Helmholtz coil pair creates a static magnetic field along the pump beam axis. The magnitude of the field is adjustable between 0 and 0.54\,G, corresponding to Larmor frequencies up to 250\,kHz~\cite{RF1,RF2}. The current source for the magnetic field is comprised by two laser diode current drivers (Thorlabs ITC 502) connected in parallel and controlled via their modulation inputs.
The modulation input is supplied by a GPIB-controlled voltage source (Krohn-Hite Model 523). 

An oscillating magnetic field at the Larmor frequency causes the spins to precess about the leading field axis. The resulting spin precession is read out via the polarization rotation of a linear polarized probe laser beam \cite{Szymek1} (Toptica DL Pro, 780\,nm, 20\,mW, $\diameter\approx5$\,mm). A small fraction of the probe beam is passed twice through a 350\,MHz acousto-optical modulator (AOM) for the probe-laser frequency stabilization. This component of the beam is, as a result, 700\,MHz lower in frequency and then used to generate a DF-DAVLL error signal with a separate, uncoated vapor cell. The laser is locked to the D2 resulting $F=3 \rightarrow F'=3,4$ crossover feature in this cell so that the probe's laser light is approximately +640\,MHz higher in frequency than the $F=3 \rightarrow F'=4$ transition. 

The probe beam traverses the cell orthogonal to the pump beam and its polarization change is measured by a balanced polarimeter comprised of a polarizing beamsplitter cube and a balanced photodetector. (Thorlabs PDB210A). The oscillating signal is analyzed with a lock-in amplifier (Signal Recovery 7265). The lock-in amplifier frequency range of 250\,kHz is the limiting factor for the frequencies used in this experiment. 

The rf driving coil is a custom-made, five-turn solenoid with an inner diameter of 0.5\,mm and a copper wire thickness of 0.2\,mm. The calculated inductance of the coil is on the order of a few nH and the coil behaves primarily as a resistive load within the frequency range reported here. It is positioned in the plane spanned by the pump and probe beams, roughly 5\,cm away from the cell center  along the static magnetic field axis. The orientation of the solenoid axis is perpendicular to the pump and probe beam. It is supplied with an oscillating current via a 20\,cm rigid coaxial conductor pair. The outer conductor is a 5\,mm diameter hollow aluminium pipe with 0.5\,mm wall thickness to prevent the emission of spurious rf from the current leads. The current is driven by the amplified output of a computer controlled function generator (SRS DS335, amplifier AE Techron 7224-P). 
An applied signal at 10\,kHz with amplitude 1\,$\text{V}_{\text{pp}}$ creates an oscillating magnetic field with approximately 5\,G amplitude just below the solenoid. The calibration for frequencies up to 10\,kHz was done with a commercial gaussmeter. 
The oscillating return flux of the solenoid is detected by the magnetometer if the driving frequency matches the Larmor frequency of the atoms in the cell.

\begin{figure}
  \centering
  \includegraphics[width=\columnwidth]{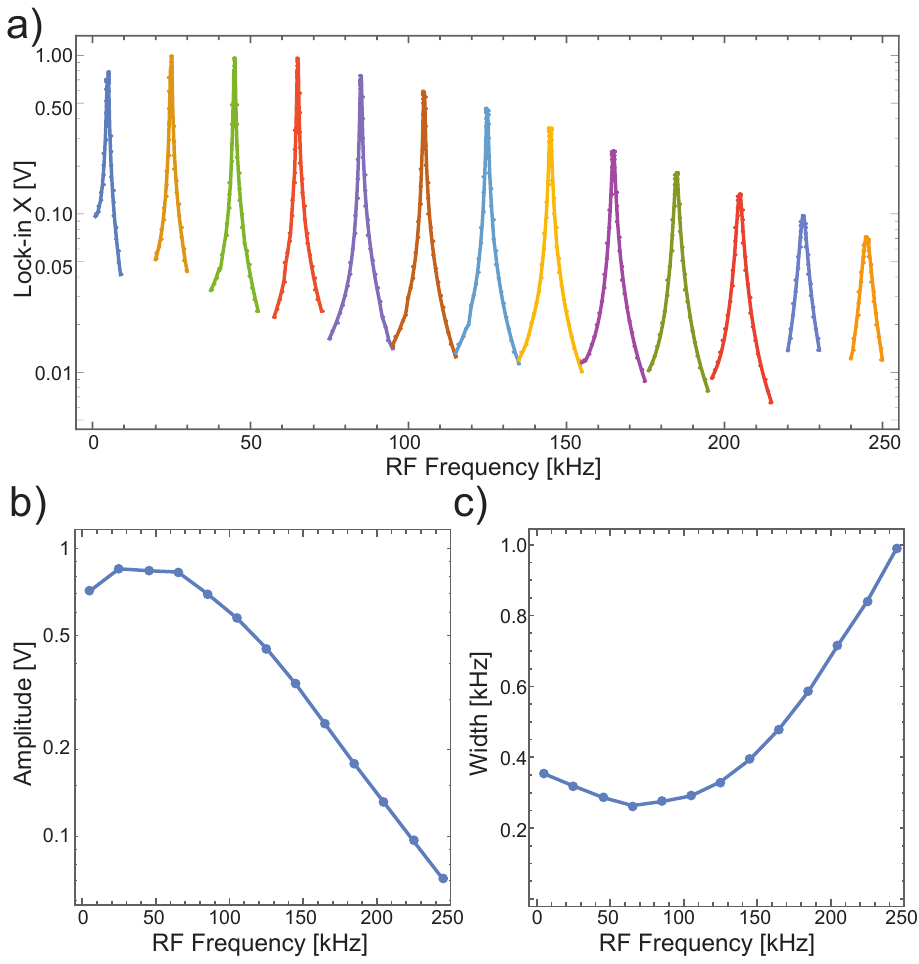}
  \caption{a) Magnetic resonances of the \rubidium cell for different applied magnetic fields.
b) Amplitude of the frequency response.
c) Width of the resulting Lorenzian line shapes.
The amplitude of the magnetic resonance goes down for two reasons: the applied voltage over the coil is kept constant and its low-pass characteristic reduces the field; the width of the resonance goes up due to uncompensated inhomogeneities of the background field.}\label{Figure2}
\end{figure}

In order to demonstrate the magnetometer's capacity to operate at different radio-frequencies, we performed measurements of the magnetic resonance (MR) for different static magnetic fields. Figure \ref{Figure2} shows a selection of forced-oscillation scans and their analysis.
In a forced-oscillation scan, the static-field current is set to a given value and the driving frequency is scanned over the Larmor frequency. At each frequency the amplitude and the phase of the oscillating polarization signal is recorded via the lock-in amplifier. The resulting data in Fig.~\ref{Figure2}a) show the Lorentzian-shaped magnetic resonances for several different fields. The measured Larmor frequencies, $\omega_L$, the amplitudes and the widths of the resonances can be extracted from nonlinear least-squares fits using the predicted lineshapes. The amplitudes and widths of the MR are displayed in Fig.~\ref{Figure2}b) and c) respectively. The reduction in amplitude can be attributed to two technical features. First, the gain of the amplifier reduces starting around 100\,kHz, so the actual applied rf field amplitude decreases with frequency. Second, the magnetic resonance increases in width mostly due to magnetic field gradients within the cell volume, which are a function of the leading field strength. This increase in resonance width reduces the amplitude further. These limitations can be lifted by using an amplifier optimized for higher frequencies and by applying compensation gradients, respectively. 

The experimental procedure for acquiring eddy-current images is as follows:
First, a sample is placed on a non-conductive mount that is attached to a three-axis computer-controlled translation stage and positioned away from the driving coil. 
Second, the rf is tuned to the center of the magnetic resonance feature for a given leading magnetic field and the lock-in reference phase is set to zero when there is no material. 
Finally, the sample is moved with the translation stage. At each position of a scan, the phase and amplitude of the lock-in signal is recorded together with the coordinates of the translation stage.

\section{\label{sec4}Results and analysis}

\begin{figure}
  \centering
  \includegraphics[width=\columnwidth]{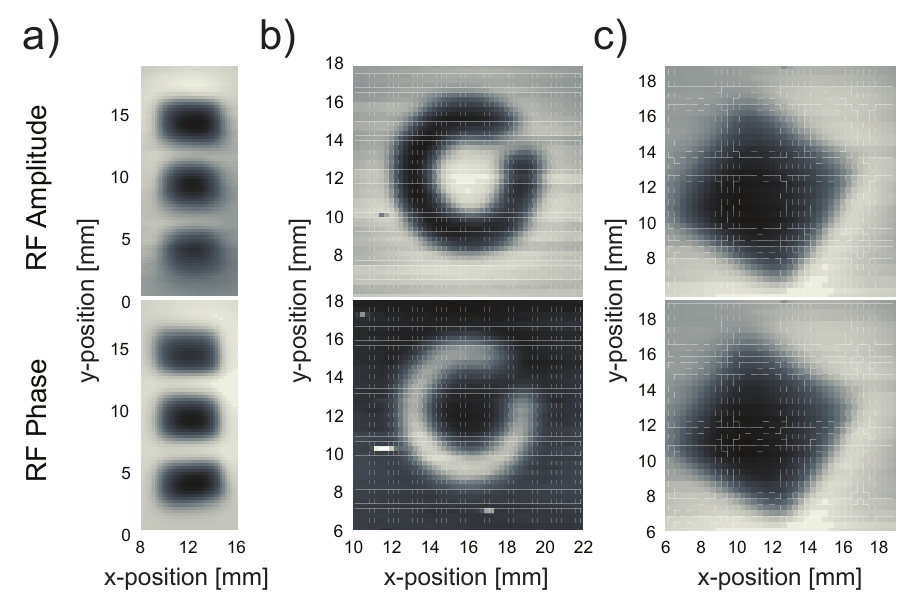}
  \caption{Images of raw lock-in data for different metallic objects, demonstrating sub-mm imaging resolution. a) Top to bottom: copper, aluminium, brass rectangles imaged with $\omega_{\text{RF}}/2\pi=80$\,kHz, and object dimensions $\left(5,3.5,0.5\right)$\,$\text{mm}^3$. b) Copper ring with 1\,mm cut, $\omega_{\text{RF}}/2\pi=150$\,kHz. The disturbances in the phase image at $\left(x,y\right)=\left(12,10\right)$ are a signature for transient magnetic field changes due to activity in the lab. c) Copper square with dimensions $\left(7.5,7.5,1\right)$\,$\text{mm}^3$ imaged with $\omega_{\text{RF}}/2\pi=80$\,kHz,}\label{Figure0}
\end{figure}

As a demonstration, Fig. \ref{Figure0} shows images of selected conductive objects at 80 and 150\,kHz. All the images show sub-mm resolution and high contrast in the amplitude and phase of the detected rf signal (raw data). The images are displayed in grey scale to avoid misinterpretation of the objects borders due to false coloring. Figure \ref{Figure0}a) shows three different materials imaged in the same scan. The contrast of the phase and amplitude plots is material-dependent, which will be discussed in more detail below. Fig. \ref{Figure0}b) shows an image of a copper ring featuring a 1\,mm cut to provide an example relevant to the important industrial application of crack detection. 
And Fig. \ref{Figure0}c) present a copper square placed at an angle to the scan axis, to demonstrate orientation-independent sub-mm features.

\begin{figure*}
  \centering
  \includegraphics[width=\textwidth]{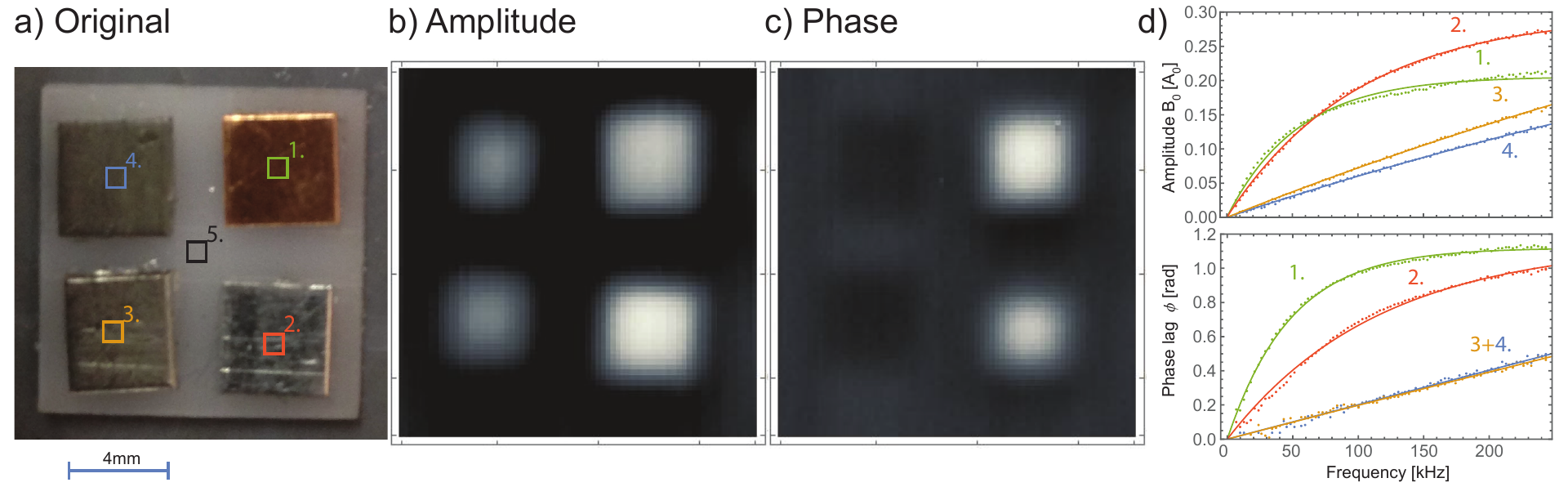}
  \caption{Magnetic induction tomography images and analysis. a) Sample holder with four different $4\times 4 \times 0.5$\,mm$^3$ slabs of metals with different conductivities. 1. copper, 2. aluminium, 3. cupronickel, 4. titanium b) Amplitude image at 245\,kHz c) corresponding phase image. d) frequency scan of the driving coil, for this just the points indicated in a) are measured and then the frequency changed. Material phase lag and response field are constructed as described in the text. Both datasets are fitted with an exponential function. This scaling could be reproduced with finite-element simulations.}\label{Figure3}
\end{figure*}

In a second experiment we wanted to demonstrate how the broad frequency range can be used to discriminate different objects according to their conductivity. We study a set of samples made of different metals. These were $4\times 4 \times 0.5$\,$\text{mm}^3$ samples of copper, aluminum, titanium, and cupronickel, a copper-nickel compound. The metals were chosen due to their range of conductivities (titanium: $2.38\times10^6$\,S/m,  cupronickel:$\left(2-5\right)\times10^6$\,S/m, aluminium: $3.50\times10^7$\,S/m, copper: $5.96\times10^7$\,S/m), their lack of magnetization to avoid interference with the magnetometer. A photograph of the sample holder is shown in Fig.~\ref{Figure3}a).
A magnetic induction image was taken of the four samples at a driving frequency of 245\,kHz, with 200\,$\mu$m step sizes and a lock-in time constant of 100\,ms. The resulting images created from the measured lock-in values can be seen in Fig.~\ref{Figure3}b) and c). All the squares are visible in the phase and amplitude data with sub-millimeter resolution.

As a final experiment we performed a frequency scan at fixed points to discriminate the different materials. To understand the resulting data, we briefly review how the induced magnetic field is extracted from the lock-in data.
The signal due to the oscillating magnetic field, $M\left(t\right)$, measured at the position of the vapor cell has two components: one from the driving field (and potentially leaking fields from the wires delivering the current to the driving coil) $A\left(t\right)=A_0 \sin\left(\omega t\right)$ and another component due to the induced eddy currents $I_{EC}\left(t\right)$. In the quasi-static approximation the induced magnetic field component is proportional to and in-phase with the oscillating eddy current $B\left(t\right)\propto I_{EC}\left(t\right)$ (see for example \cite{sensors2011}).The signal detected with the lock-in amplifier can thus be written as
\begin{equation}
M_M\sin\left(\omega t+\psi_M\right)= A_0 \sin\left(\omega t\right)+ B_0 \cos\left(\omega t-\phi\right),
\end{equation}
where $M_M$ and $\psi_M$ are the measured lock-in amplitude and the phase, respectively. The lock-in phase is zeroed ($\psi_M=0$) when there is no material (position~5 in Fig.~\ref{Figure3}a). In the absence of material the measured amplitude is just the amplitude of the driving field at the position of the cell $M_M=A_0$. The induced field amplitude, $B_0$, and its phase shift, $\phi$, can then be related to the measured values by rearrangement of Eq.~1:
\begin{align}
\phi&=\arctan\left(\frac{A_0-M_M\cos\left(\psi_M\right)}{M_M\sin\left(\psi_M\right)}\right),\\
B_0&=\sqrt{M_M^2\sin\left(\psi_M\right)^2+\left(A_0-M_M\cos\left(\psi_M\right)\right)^2}.
\end{align}

For each point in Fig. \ref{Figure3}d) the magnetic field was set via GPIB and a forced oscillation scan was performed. The resulting data were fit with a Lorentzian to determine the Larmor frequency. The frequency of the drive was set to the Larmor frequency and the lock-in phase zeroed. Four points on top of the different materials (position 1-4) were measured and five points at positions without material (summarized as position 5 in Fig. \ref{Figure3}a). The mean of the five points without material were used to determine $A_0$. In total, magnetic induction measurements were performed for 87 different frequencies ranging from 7\,kHz to 243\,kHz with an average step size of 2.75\,kHz.

The amplitude of the signal due to the response field, $B_0$, and the phase lag $\phi$ as a function of the frequency can be seen in Fig.~\ref{Figure3}d). All data were normalized with $A_0$ and the phases were shifted by the same constant for all the data points such that all phase curves pass through the origin.

All the metals are clearly distinguishable. The response field amplitude and the phase $\phi$ increase linearly for low frequencies with a slope related to the conductivity. At high frequencies, the two metals with the highest conductivity (copper and aluminum) saturate, which could be explained by skin depth effects and can actually be used to measure the thickness of the material \cite{Multifrequency1}. The data are well-approximated by an exponential function, and this behavior has been confirmed by fine-element models. The skin-depths calculated from the decay-constant frequencies and known conductivities approximately correspond to the materials' thickness, however the exact quantitative relationship between these is the topic of future work.  

\section{\label{sec5}Summary and outlook}
We have presented sub-mm resolution eddy current images with an atomic radio-frequency magnetometer for different frequencies and materials with different conductivities. In a first set of measurements we demonstrated the feasibility of the device to detect amplitude and phase of fields from induced eddy currents for frequencies between 7-250\,kHz, which is technically limited only by the frequency range of the lock-in amplifier and not by the atomic sensor. Extending the operating frequency into the MHz range is a straightforward technical improvement that can open the possibility for biological applications.

In a second set of measurements we demonstrate how the frequency scanning capacity of the alkali-based rf magnetometer can be used to discriminate different materials with varying conductivity. 
Eddy current detection is commercially widely used but mostly done with detection coils, but also with giant magnetoresistance (GMR) based magnetic field sensors and super-conducting quantum interference devices. We add alkali-based rf magnetometer to the spectrum which might have substantial benefits for low-frequency and therefore high-penetration depth eddy-current detection, due to their much higher sensitivity than rf coils and their miniaturization capacity. 
Another extension of this method currently being pursued involves replacing the magnetometer in this setup with one based on nitrogen-vacancy centers in diamond. This would be beneficial for spatial resolution, scalability and eddy-current detection in the GHz range. 

We acknowledge support by the DFG through the DIP program (FO 703/2-1). NL was supported by a Marie Curie International Incoming Fellowship within the 7th European Community Framework Programme.
\bibliography{literature}
\end{document}